\documentclass[showkeys,notitlepage,twocolumn]{revtex4-1}%
\usepackage{amsmath,amssymb,amsbsy,booktabs,subfigure,hyperref,subfigure,bbm,times,txfonts,accents,float}
\usepackage{caption}
\usepackage{graphicx}
\usepackage{dcolumn}
\usepackage{bm}
\usepackage{lineno,hyperref}
\usepackage{graphicx}
\usepackage{epstopdf}
\modulolinenumbers[5]
\emergencystretch=\maxdimen
\hypersetup{colorlinks=true,linkcolor=blue,anchorcolor=blue,citecolor=blue}

\begin{document}

\title{Neural Network-Based Multipartite Entanglement Classification with Prior Guidance from Quantum Uncertainty Relations}

\author{Qiyi Li}
\author{Xiao Zheng}
\email{xiaozheng@buaa.edu.cn}
\author{Guofeng Zhang}
\email{gf1978zhang@buaa.edu.cn}

\affiliation{%
	School of Physics, Beihang University, Xueyuan Road No. 37, Beijing 100191, China
}%

\date{\today}

\begin{abstract}
Quantum entanglement is a crucial resource in quantum information processing, yet its efficient, scalable and robust classification in multipartite systems remains theoretically challenging. Although supervised machine-learning has been applied to this task, most existing methods still suffer from high measurement costs, computational consumption, and weak noise robustness. In this work, by incorporating multipartite uncertainty relations as prior guidance, we propose a neural network approach to classify distinct Stochastic Local Operations and Classical Communication (SLOCC) multipartite entanglement classes based on states sampled from their local unitary (LU) orbits. Compared with traditional techniques, our method reduces experimental measurement-resource requirements and computational overhead, showing high adaptability to large-scale systems. The classification accuracy of our method reaches 99.5\% in 20-qubit systems. The numerical validation is performed on states generated by random LU transformations, which preserve the SLOCC class. Within this setting, the proposed method offers strong effectiveness, scalability, and robustness for multipartite entanglement classification.
\end{abstract}


\maketitle

\section{Introduction}
Quantum entanglement\cite{I.1,I.2,I.3}, which is considered as a fundamental feature and advantage of quantum systems compared to classical systems, has been widely applied to quantum information\cite{I.4,I.5,I.7,I.8}, quantum computation\cite{I.11,I.12,I.13,I.new.computation1}, quantum cryptography\cite{I.14,I.new.cryptography1,I.new.cryptography2} and quantum metrology\cite{I.15,I.16}. The case of entanglement in bipartite systems is relatively simple and well studied as there is essentially only one class of entanglement between two qubits. Nevertheless, it is far more complex and challenging for multipartite systems in which the qubits can be entangled in various inequivalent ways\cite{I.50}. The different classes of entangled states can be distinguished by their unique properties and behaviors under Stochastic Local Operations and Classical Communication (SLOCC). Under SLOCC, two multipartite states belong to the same class of entanglement if and only if they can be transformed into each other. Accordingly, multipartite states can be distinguished into different classes and each of them is studied for different specific application scenarios\cite{I.16,I.18,I.19}. Thus, low-cost and efficient methods to distinguish the entanglement properties of quantum systems are highly desirable and have gained widespread attention in recent years\cite{I.20,I.21,I.22,I.23}.\\

For decades, a multitude of analytical approaches have been proposed for distinguishing the entanglement properties of quantum systems. Among these methodologies, entanglement witnesses\cite{I.25,I.26,I.28,I.29} and quantum state tomography\cite{I.30,I.31,I.32} are most commonly used, both of which rely on semidefinite programming (SDP) as their fundamental algorithmic framework. However, the computational complexity of these techniques scales exponentially with the dimensionality and the number of the subsystems, which demonstrates that the underlying problem is non-deterministic polynomial-time hard for large-scale quantum systems\cite{I.32.A}. To address these challenges, a series of novel analytical techniques have been proposed\cite{I.32.B,I.32.C,I.32.D}. These approaches introduce a series of polynomial entanglement quantifiers and use local measurements to evaluate them, which significantly reduces the computational complexity. In this way, the detection of states with genuine multipartite entanglement (GME) in a 13-qubit system has been successfully demonstrated\cite{I.32.B}. Nevertheless, these analytical methods focus primarily on the pure states, and the detection of entangled mixed states remains a key challenge for these approaches\cite{I.32.B,I.32.C,I.32.D}. Hence, a universally applicable method for characterizing entanglement in both pure and mixed states is critically needed.\\

In recent years, machine learning approaches, as more efficient entanglement classification techniques, have attracted extensive research attention\cite{I.33,I.34,I.23.1}. Compared with traditional techniques, machine learning approaches are capable of reducing computational complexity from exponential growth to polynomial scaling, owing to their strong ability to extract inherent features of quantum systems\cite{I.23}. Furthermore, these methods are applicable to a wider range of tasks, including the classification of both pure and mixed quantum states. Initially, supervised learning methods were adopted\cite{I.22,I.33,I.34}. Leveraging convolutional neural networks (CNNs) or support vector machines (SVMs), these studies have achieved high accuracy for specific entanglement classification tasks. Nevertheless, most existing supervised-learning-based entanglement classification approaches are still plagued by several critical limitations. First, the majority of existing methods adopt the full density matrix as the input feature, which requires massive quantum measurements in practical scenarios. This limitation originates from the bottlenecks of current quantum state tomography, particularly for large-scale quantum systems exceeding 12 qubits, further exacerbating the inefficiency in the preparation of labeled data\cite{I.22}. Although certain study has extended such methods to 20-qubit systems, this work is merely confined to the binary classification between GME states and separable states, thereby lacking universal applicability\cite{I.23}. Second, the computational consumption for training these methods increases rapidly with the number of qubits because the dimension of the density matrix grows exponentially with both subsystem dimensionality and subsystem count, which severely restricts their scalability to large quantum systems. Third, most of these methods primarily focus on computational and mathematical aspects, with insufficient investigation into the physical characteristics of quantum systems.\\

The uncertainty relation is one of the most fundamental principles in quantum mechanics and reveals the inherent limitations on the precision of simultaneously measuring incompatible observables\cite{I.36,I.37,I.38,I.39,I.40,I.41,I.42}. For decades, the uncertainty relation has been regarded as a reliable technique for detecting the entanglement properties in bipartite quantum systems as the lower bound of uncertainty can be reduced by quantum entanglement\cite{I.43.A,I.43.B}. Recently, a variance-based quantum-control-assisted multipartite uncertainty relation has been established. It demonstrates that the lower bounds of the corresponding uncertainty relation for states within specific entanglement structures can be violated to different extents\cite{I.48}. That is to say, the uncertainty of incompatible observables is closely related to quantum entanglement and can serve as a powerful tool for distinguishing multipartite entanglement properties.\\

Therefore, in this work, we incorporate a recently-constructed variance-based multipartite quantum-control-assisted uncertainty relation into the input feature extraction for machine learning models, thereby proposing a novel neural network-based entanglement classification method for multipartite quantum systems, which is guided by the prior knowledge derived from quantum uncertainty relations. By introducing this novel input feature, our method effectively reduces the measurement resource consumption in practical experiments and the computational burden in machine learning processes, demonstrating high adaptability to large-scale multipartite quantum systems and exceptional reliability under noisy conditions. The proposed method offers sufficient effectiveness, scalability and robustness for both pure states and mixed states, and demonstrates satisfactory performance in ternary classification between GHZ-class, W-class and linear cluster-class states in the quantum system with the number of qubits spanning from 3 to 20.\\ 

The remainder of this paper is structured as follows. Section II presents the framework of the proposed method and the details of input features extraction. Section III elaborates on the methodologies for dataset construction. Section IV presents the numerical results and performance evaluations of the proposed approach for distinguishing entanglement classes in multipartite quantum systems. Finally, we summarize the core findings and contributions of this work in Section V.\\

\section{Methodology: Uncertainty-Guided Feature Extraction}
The variance-based quantum-control-assisted multipartite uncertainty relation reads:\cite{I.48}\\

\begin{align} 
    \tag{1}
    \sum_{k=1}^{K}E\left [ V(Q_{k}^{A}|O_{k}^{C_{1}},\cdot \cdot \cdot ,O_{k}^{C_{n}}) \right ]\ge L_{tra}-\sum_{k=1}^{K}V\left [ E(Q_{k}^{A}|O_{k}^{C_{1}}) \right ] \nonumber \\
    -\sum_{k=1}^{K} \sum_{n=2}^{N}E\left [ V(E\left [ Q_{k}^{A}|O_{k}^{C_{n}} \right ]|O_{k}^{C_{1}},\cdot \cdot \cdot ,O_{k}^{C_{n-1}}) \right ] \nonumber
\end{align}
where $O_{k}^{C_{n}}$ and $Q_{k}^{A}$ represent the carefully selected observables $O_{k}$ and $Q_{k}$ performed on particle $C_{n}$ and $A$, respectively. Particles $C_{1},\cdot \cdot \cdot ,C_{n}$ denote the control subsystems while particle $A$ refers to the measured subsystem. $E\left [ V(Q_{k}^{A}|O_{k}^{C_{1}},\cdot \cdot \cdot ,O_{k}^{C_{n}}) \right ]$ is the conditional variance of $Q_{k}^{A}$ when one has performed the measurements $O_{k}$ on particles $C_{1},\cdot \cdot \cdot ,C_{n}$. $V\left [ E(Q_{k}^{A}|O_{k}^{C_{1}}) \right ] $ is the variance of the conditional expectation of $Q_{k}^{A}$ under the condition that the measurement $O_{k}^{C_{1}}$ has been performed. $E\left [ V(E\left [ Q_{k}^{A}|O_{k}^{C_{n}} \right ]|O_{k}^{C_{1}},\cdot \cdot \cdot ,O_{k}^{C_{n-1}}) \right ] $ is defined as the conditional variance of $E(Q_{k}^{A}|O_{k}^{C_{n}})$ when one has performed the measurements $O_{k}^{C_{1}},\cdot \cdot \cdot ,O_{k}^{C_{n-1}}$. Finally, the term $L_{tra}$ stands for the traditional lower bound for separable states.\\

To incorporate relation (1) into the preparation of input features for machine learning, a brief introduction to the uncertainty relation is necessary. The physical interpretation of relation (1) can be illustrated by the following game, shown in the left panel of Fig.1(b)\cite{I.48}: i. Bob prepares a multipartite quantum system, which includes a measured subsystem $A$ and several control subsystems $C_{1},C_{2},\cdot \cdot \cdot ,C_{n}$. ii. Bob sends the measured subsystem $A$ to Alice. iii. Alice selects a measurement to be taken, such as $Q_{k}$, and informs Bob of her choice. iv. Based on the available information about the quantum state of the whole system, Bob chooses an appropriate measurement, such as $O_{k}$, and performs $O_{k}$ on the control subsystems $C_{1},C_{2},\cdot \cdot \cdot ,C_{n}$. v. Alice performs $Q_{k}$ on the measured subsystem $A$. The measurements performed on control subsystems $C_{1},C_{2},\cdot \cdot \cdot ,C_{n}$ can be used to control the whole system to achieve the minimal uncertainty of the measured subsystem $A$. Therefore, the measurements performed on the corresponding control subsystems are called quantum control. In this way, the traditional lower bound can be broken.\\

The formulation of relation (1) provides a powerful tool for investigating the uncertainty in multipartite quantum systems. As proven by prior research\cite{I.46,I.47,I.48}, the uncertainty of incompatible measurements inherently encodes critical information about quantum entanglement within the system. In other words, the conditional uncertainty can potentially be employed as an input feature for machine learning-based entanglement classification. Compared with the full density matrix adopted by most existing machine learning approaches, which has a dimensionality of $O(n)_{1} = d^{n} \times d^{n}$, the conditional uncertainty of the system has a dimensionality of $O(n)_{2} = d \times n$, where $n$ denotes the number of subsystems and $d$ is the dimension of each subsystem. Obviously, the latter requires much fewer measurement resources in practice and entails lower computational costs in machine learning. Precisely because our method employs the conditional uncertainty of the quantum system rather than the full density matrix as the input features for the neural network, it offers the following advantages:\\

1. In the input features preparation, the measurement resources required to obtain the full density matrix of a quantum system are far larger than that required to obtain its conditional uncertainty. For this reason, compared with existing mainstream methods, our approach substantially improves the efficiency of the aforementioned preparation and consequently accelerates model training.\\

2. In the model training, the dimension of the conditional uncertainty of the quantum system scales polynomially, rather than exponentially, with the number and dimension of subsystems, which makes our method particularly well-suited for large-scale multipartite systems to which existing mainstream methods are difficult to scale due to computational intractability.\\

3. Finally, compared with the full density matrix, the conditional uncertainty better captures the essential entanglement characteristics of the quantum system, making our method more robust than mainstream approaches in both mixed-state and noisy scenarios. Furthermore, the conditional uncertainty has a clear physical meaning. Employing it as an input feature of the neural network enables us to break out of the "AI black box", allowing us to investigate this issue from a perspective that is closer to the physical essence.\\

Here, a comparison between our method and other existing mainstream entanglement classification methods based on machine learning is made, as shown in Table I.\\

\begin{table*}[!t]
    \centering
    \caption{Comparison between our method and other existing mainstream entanglement classification methods based on machine learning. Input Features denotes the physical nature of the training data; Growth of complexity indicates whether the cost of generating training samples scales polynomially or exponentially with the number of qubits; the last three columns report the largest number of qubits tested and the corresponding classification accuracy (Acc) on pure, mixed, and noisy quantum states, respectively. / means that the original work did not provide a separate evaluation under that specific scenario.}
    \begin{tabular}{@{}lcccccc@{}}
        \toprule[0.5pt]
        Cited form & Year & Input Features & Growth of complexity & Pure states & Mixed states & Noisy states \\
        \midrule[1pt]
        This work & 2026 & Conditional uncertainty & Polynomial & 20 qubits, Acc~(0.995) & 10 qubits, Acc~(0.985) & 10 qubits, Acc~(0.993) \\
        \addlinespace
        Ref[20] & 2023 & Feature vectors & Polynomial & / & 2 qubits, Acc~(0.932) & 10 qubits, Acc~(0.984) \\
        \addlinespace
        Ref[21] & 2023 & Feature vectors & Exponential & / & 3 qubits, Acc~(0.905) & / \\
        \addlinespace
        Ref[22] & 2024 & Full density matrices & Exponential & / & 11 qubits, Acc~(1.000) & 10 qubits, Acc~(0.983) \\
        \addlinespace
        Ref[37] & 2025 & Full density matrices & Exponential & 4 qubits, Acc~(0.872) & 4 qubits, Acc~(0.659) & / \\
        \bottomrule[0.5pt]
    \end{tabular}\label{TableI}
\end{table*}

Based on the above analysis, a new neural network-based method to detect the entanglement classes of multipartite quantum systems with the prior guidance from uncertainty relations is constructed. A schematic diagram of the proposed method is shown in Fig.1. Briefly speaking, it can be divided into three parts, i.e., the state preparation, the feature extraction, and the model training, which are demonstrated in Fig.1(a), Fig.1(b) and Fig.1(c), respectively.\\

\begin{figure*}[!t]
    \centering
    \resizebox{16cm}{7.5cm}{\includegraphics{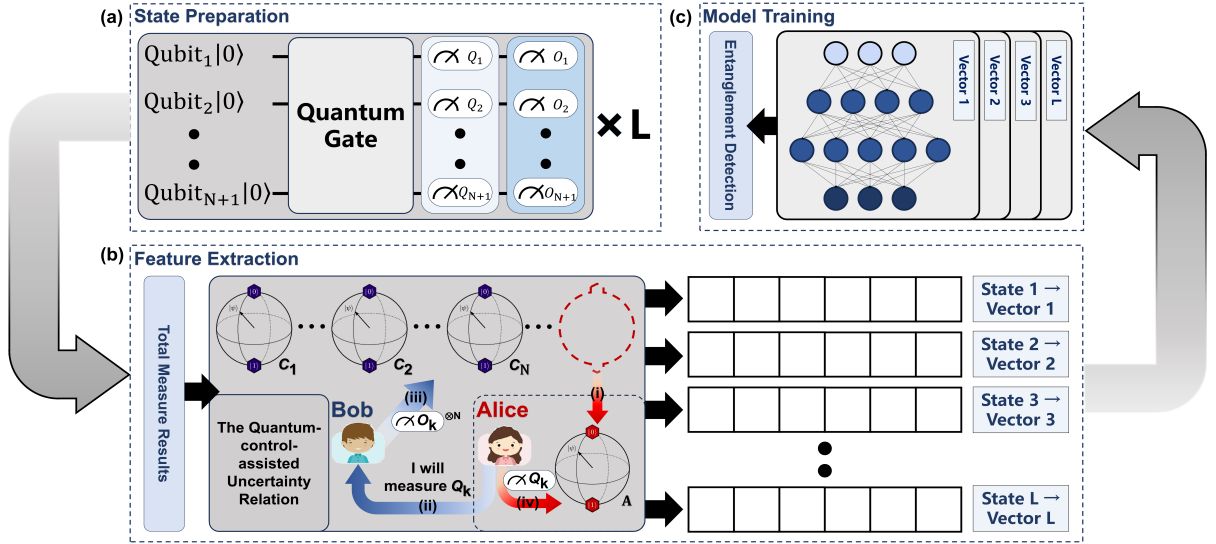}}
    \caption*{Fig.1 Schematic diagram of the proposed method. In the first step, i.e., the state preparation, a total of L random quantum states are generated and measured, thereby providing the necessary data for the following steps. In the feature-extraction step, we calculate the conditional uncertainty for each qubit based on the measurement results and the uncertainty relation. Subsequently, the calculated results are considered as the input features and used for model training.}\label{Fig1}
\end{figure*}

For the state preparation, we generate a large number of random quantum states with specific entanglement classes, including both pure and mixed states, and perform measurements of carefully selected incompatible local observables over a large number of experimental runs. The detailed generation and measurement procedures will be described in the next section.\\ 

Using the measurement results, the input features for machine learning can be prepared. According to Fig.1(b) and relation (1), the left-hand side of relation (1) represents the conditional uncertainty of the system, expressed in the form of conditional variance. In certain prior studies, the conditional uncertainty of the system has exhibited distinct spectral signatures for different multipartite entanglement structures, thus serving as an effective feature for multipartite entanglement classification\cite{I.48}. Therefore, we can calculate the conditional uncertainty for each qubit on the basis of relation (1) and the aforementioned measurement results. The derived conditional uncertainty values corresponding to each qubit in an n-qubit system collectively constitute an n-dimensional vector, which can be utilized as the input feature vector for the proposed machine learning method.\\

Finally, for the model training, we adopt a standard classical multilayer perceptron (MLP) as our neural network architecture, and train it on the prepared dataset to classify the entanglement classes of multipartite quantum systems. The detailed parameters of the MLP architecture will be discussed in the following section.\\

\section{The Generation of Dataset}
Nowadays, different application scenarios demand various classes of entangled states. Thus, the classification of these states is crucial for their effective utilization, especially in the case of GME states. Among the GME states, Greenberger-Horne-Zeilinger (GHZ) states\cite{I.49}, W states\cite{I.50} and linear cluster states\cite{I.51} are relatively common and important\cite{I.12,I.16,I.18,I.19}. Thus, this work focuses primarily on the classification of these three SLOCC-distinct entanglement classes. In this work, we adopt the quantum circuit method to construct our dataset for more efficient preparation of quantum states and lower consumption of measurement resources, leveraging the recently released Origin Quantum platform\cite{I.52}.\\

The corresponding quantum circuits for GHZ states and linear cluster states are illustrated in the upper panel of Fig.2. Specifically, for GHZ state preparation, we first apply a Hadamard gate to the first qubit to generate superposition, which is followed by a series of CNOT gates to entangle all subsequent qubits. For linear cluster states, by contrast, we apply Hadamard gates to all qubits to create superposition first, and then implement controlled-Z (CZ) gates between neighboring qubits to establish linear chain entanglement. As for W states, we initially define the amplitudes of the basis vectors and then adopt the approximate matrix product state (MPS) encoding method provided by Origin Quantum to generate the target states. The detailed construction process is shown in the lower panel of Fig.2. After preparing these specific quantum states, random entangled states belonging to specific SLOCC classes can be generated. For each prepared state, we subsequently apply a series of random local unitary (LU) one-qubit gates $U_{1} \otimes U_{2} \otimes \cdots \otimes U_{n}$ on the qubits, where each $U_{i}$ refers to a randomly chosen single-qubit unitary acting on the $i$-th qubit. This step is referred to as a random LU transformation, which can be used to generate a variety of entangled states without altering their SLOCC entanglement classes since local unitaries form a subgroup of the SLOCC operations\cite{I.50}. That is, while the states inside each of the three generated classes are equivalent under LU transformations, any pair of states from distinct classes are inequivalent under SLOCC. Consequently, the classification target remains the SLOCC class (GHZ, W, or linear cluster), which is the physically relevant category for quantum information tasks. This clarifies why the terminology "SLOCC class" is used throughout the paper, even though the data augmentation is implemented via LU transformations. By implementing these quantum circuits on the Origin Quantum platform, we can efficiently generate, with satisfactory fidelity, a set of random multipartite entangled states with specific entanglement classes for our dataset. For the convenience of the subsequent analysis, the entanglement states transformed from GHZ state are denoted as GHZ-class here. The W-class and linear cluster-class are denoted similarly to the GHZ-class.\\

\begin{figure*}[!t]
    \centering
    \resizebox{16cm}{8cm}{\includegraphics{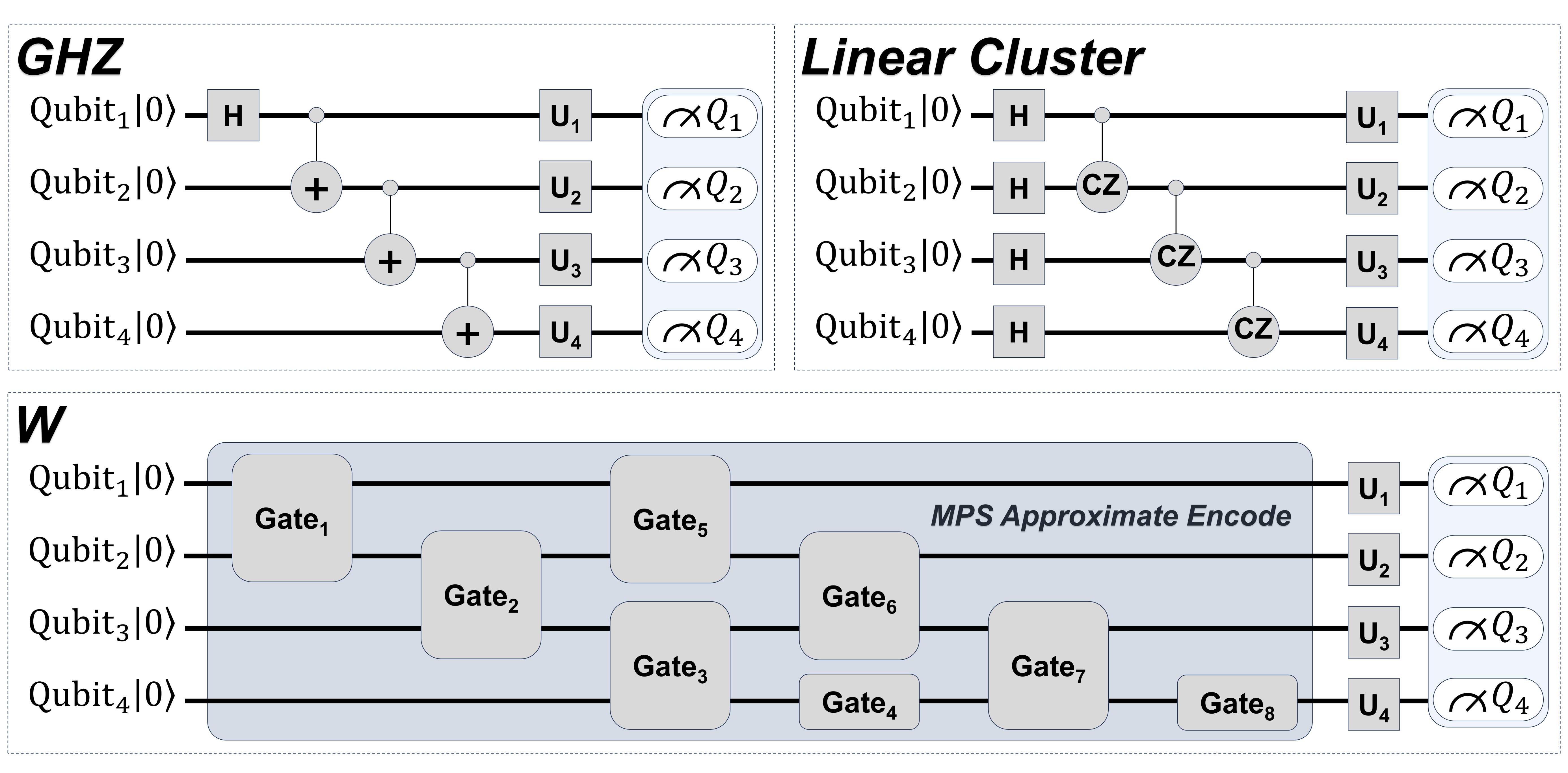}}
    \caption*{Fig.2 Schematic diagram of the quantum circuits for GHZ-class states, linear cluster-class states and W-class states. Random local unitary gates $U_{i}$ are applied independently to each qubit. Each $U_{i}$ is drawn randomly and may be different.}\label{Fig2}
\end{figure*}

However, the states generated by the aforementioned method are all pure states. In practice, mixed states have more extensive applications, and their classification is more challenging. To address this issue, we also generate mixed states by incorporating quantum ensembles. Specifically, consider a mixed state $\rho$ composed of two pure states $\left | \psi_{1} \right \rangle$ and $\left | \psi_{2} \right \rangle$, i.e., $\rho = p_{1}\left | \psi_{1} \right \rangle \left \langle \psi_{1} \right | + p_{2}\left | \psi_{2} \right \rangle \left \langle \psi_{2} \right |$, where $p_{1} + p_{2} = 1$ (with $p_1$ and $p_2$ denoting the probabilities of the pure states $\left | \psi_{1} \right \rangle$ and $\left | \psi_{2} \right \rangle$, respectively). The measurement result of the mixed state $\rho$ can be expressed as $M = p_{1}M_{1} + p_{2}M_{2}$, where $M_{1}$ ($M_{2}$) represents the measurement result of the pure state $\left | \psi_{1} \right \rangle$ ($\left | \psi_{2} \right \rangle$). In this context, a single measurement on the mixed state will yield a probabilistic outcome: $M_{1}$ with probability $p_{1}$ or $M_{2}$ with probability $p_{2}$. However, when the selected measurement is repeated many times-which is necessary for calculating the conditional uncertainty-the average measurement results converge to the weighted average $M = p_{1}M_{1} + p_{2}M_{2}$, in accordance with the statistical properties of the mixed states and the law of large numbers. In other words, measuring the mixed state is equivalent to measuring the pure states that constitute it a large number of times and taking the weighted average of the measurement results. Based on this approach, we can successfully generate mixed states.\\

Furthermore, the noise in quantum information processing is ubiquitous in practical scenarios, making it essential to take it into account. In this context, we express the noise model as\\
\begin{align} 
    \tag{2}
    \rho_{noisy} = p\rho + (1-p)\frac{I}{d} \nonumber
\end{align}
where $\rho$ is the ideal density matrix, and $I$ represents the identity matrix. $d$ refers to the dimension of the density matrix with $p$ being the fidelity. By adjusting the value of $p$, we can simulate different strengths of noise in the quantum circuits. It can be verified that the noisy state $\rho_{noisy}$ is a convex combination of the ideal state $\rho$ and the maximally mixed state $I/d$, indicating that it is a mixed state by definition. Thus, the noisy states can also be generated in the same way as mixed states.\\

Subsequently, we generate a total of 1,890,000 states, including 690,000 pure states, 400,000 mixed states, and 800,000 noisy states. The generated states span from 3-qubit to 20-qubit systems. For the measurements performed on the quantum systems, we adopt the Pauli measurements, i.e., $X$ and $Z$ measurements, which are mutually unbiased bases and widely used in quantum information processing. Using these measurements, we calculate the conditional uncertainty of each qubit for each state in an n-qubit system based on the quantum-control-assisted uncertainty relation introduced in Sec. II. By organizing these conditional uncertainties into n-dimensional vectors, we construct the input vectors for our neural network. Finally, by labeling these input vectors according to their corresponding entanglement classes, we successfully prepare a comprehensive dataset for training and testing our neural network-based multipartite entanglement classification method.\\

\section{The results and the performance}
In this section, the numerical performance of the proposed neural network-based method is presented. For this evaluation, we adopt the MLP—a fundamental neural network architecture—and assess its performance on the constructed dataset. In our study, the MLP model uses ReLU as the activation function and Adam as the optimizer with a learning rate of 0.0001, with a specific number of hidden layers corresponding to different qubit counts. We split the dataset into training and test sets with an 8:2 partition ratio, where the training set is utilized for model training and the test set for validating the model's generalization performance. In the following, prediction accuracy is selected as the primary evaluation metric, which is defined as the proportion of correctly classified samples in the test set. The model attained high classification accuracy in distinguishing GHZ-class, W-class and linear cluster-class, which verifies the effectiveness of our proposed method for multipartite entanglement classification.\\

We first focus on the ternary classification in 10-qubit systems, which serves as a validity test for the proposed method. The detailed results are presented in Fig.3, where t-distributed stochastic neighbor embedding (t-SNE) plots\cite{I.53} are employed to visualize the numerical performance of the model. Some clear and consistent conclusions can be drawn for both pure-state case and mixed-state case. It is evident that at the initial stage of training, the data points of these classes overlap and are poorly clustered, indicating that the model has not yet fully learned the intrinsic feature distinctions between these classes. In the later training epochs, the data points gradually separate and form compact clusters, which demonstrates that the model successfully captured the characteristic features of these high-dimensional conditional uncertainty vectors, enabling it to effectively distinguish between the three entanglement classes. With the progress of training, the clusters become increasingly well-separated, reflecting a continuous improvement in the model's classification performance. Actually, the accuracy of the model in the ternary classification in 10-qubit systems reaches 99.8\% for pure states and 97.3\% for mixed states, showing great potential for classifying multipartite entanglement classes.\\

\begin{figure*}[!t]
    \centering
    \resizebox{14cm}{7cm}{\includegraphics{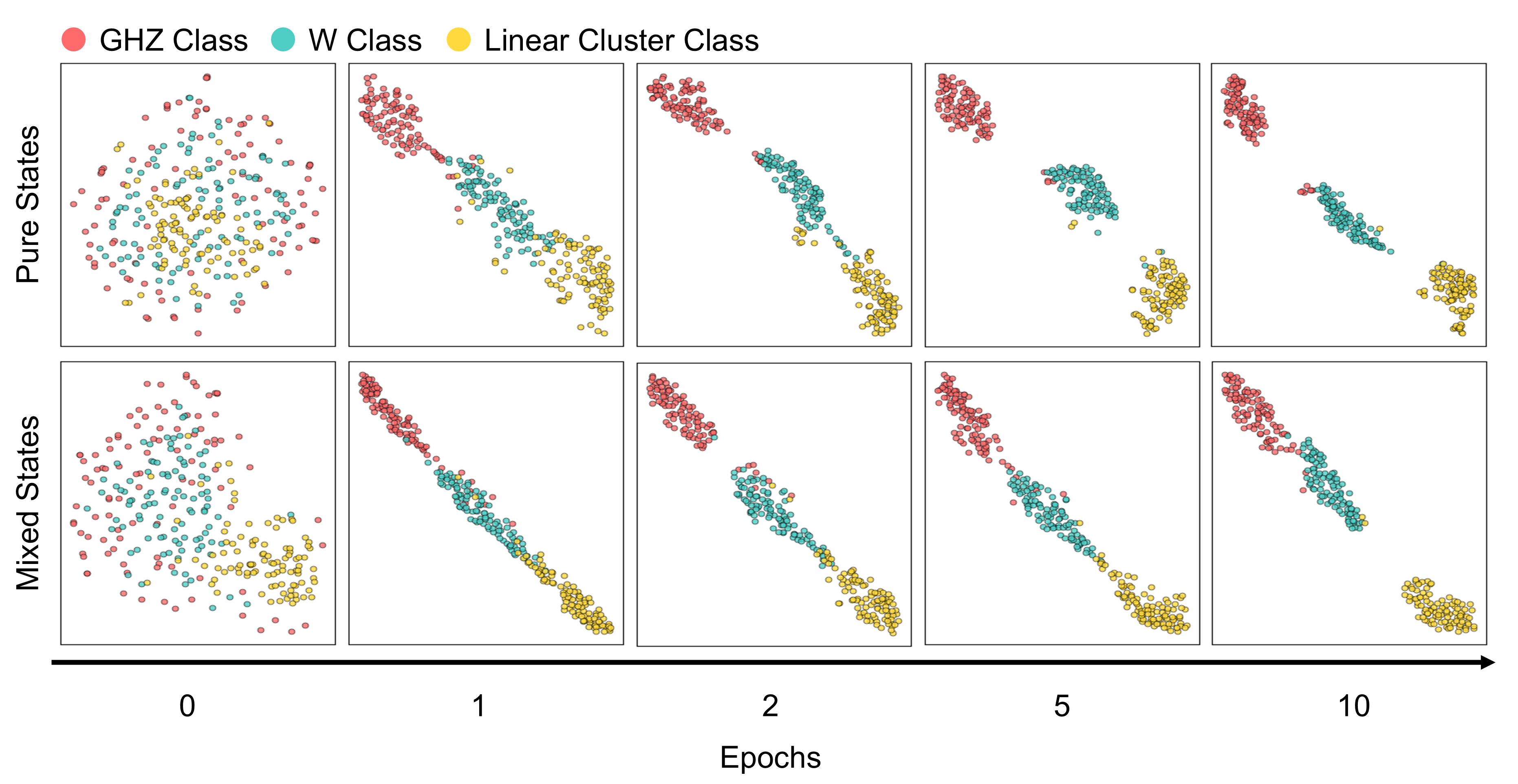}}
    \caption*{Fig.3 The t-SNE plots for the ternary classification when the number of qubits equals 10. Here a single point represents a specific quantum state. The upper panel illustrates the case for pure states while the lower panel illustrates that of mixed states, respectively.}\label{Fig3}
\end{figure*}

Building on the successful performance of our method in ternary classification, the classification of GHZ-class states and W-class states in 3-10-qubit systems is conducted. Fig.4 presents the prediction accuracy and the necessary training epochs of our method for 3-10-qubit systems with both pure and mixed states. For both pure-state case and mixed-state case, we prepared 50000 samples for each qubit number, respectively. Although the model yields a relatively low accuracy for systems with fewer qubits, the accuracy increases rapidly as the number of qubits rises. At 10 qubits, the classification accuracy for pure states reaches 99.5\%, a sufficiently high value for practical application. Meanwhile, the accuracy for mixed states is slightly lower than that for pure states but still maintains a high performance level, reaching 98.5\% for the 10-qubit mixed state system. Additionally, the number of training epochs required for convergence also remains relatively small and acceptable when the qubits number increases. That is to say, our method indeed demonstrates strong performance in medium-sized multipartite systems with both pure and mixed states.\\

\begin{figure*}[!t]
    \centering
    \resizebox{14cm}{7.5cm}{\includegraphics{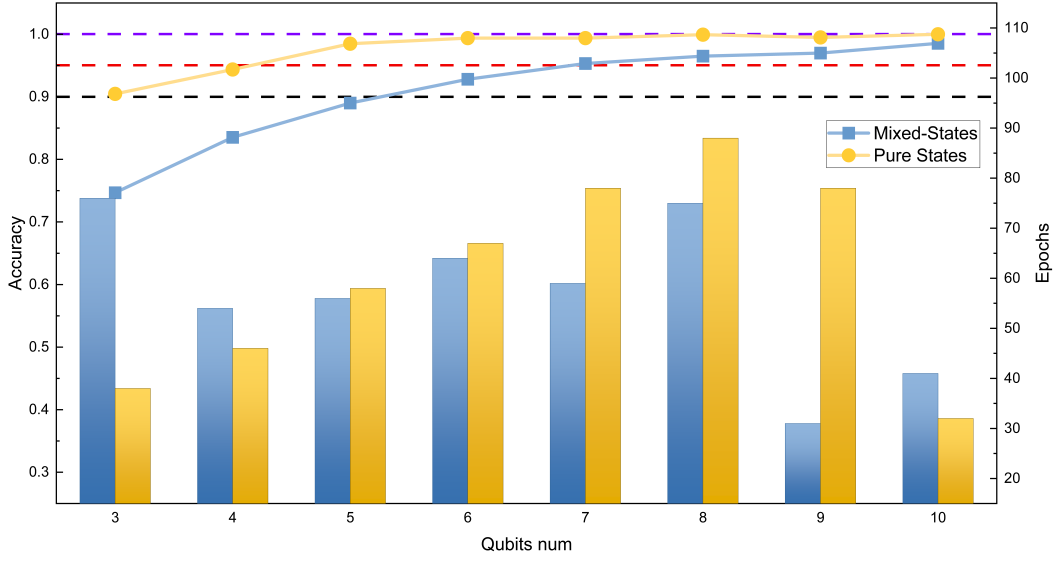}}
    \caption*{Fig.4 The numerical results for the GHZ-class and W-class classification in 3-10-qubit systems. The dot-line plot corresponds to the left vertical axis, representing the classification accuracy of entanglement classes; the histogram corresponds to the right vertical axis, denoting the minimum number of training epochs required for convergence. The black, red, and purple straight lines represent the contour line corresponding to 90\%, 95\% and 100\%, respectively.}\label{Fig4}
\end{figure*}

Specifically, we attempt to distinguish GHZ-class and linear cluster-class states across 3-20-qubit systems, as the generation of W states with satisfactory fidelity for large-scale systems is challenging. The prediction accuracy of this binary classification for pure states in 3-20-qubit systems is presented in Fig.5. When comparing the performance in 3-10-qubit systems with the previous GHZ-W binary classification, the classification accuracy shown in Fig.5 exhibits a slight decrease, especially when the number of qubits is 3. This can be attributed to the more complex physical nature of the distinction between GHZ and linear cluster states, leading to more subtle differences in their conditional uncertainty features\cite{I.48}. Nevertheless, the accuracy remains sufficiently high, reaching 99\% for 11-qubit systems. Furthermore, as the number of qubits increases, the classification accuracy for linear cluster-class states continues to rise, maintaining over 99\% for 11-20-qubit systems despite the shrinking size of the dataset. Moreover, the necessary epochs for convergence also remains relatively small and acceptable when the counts of qubits increases. This demonstrates that our method possesses strong capability in multipartite entanglement classification, and is particularly well-suited for identifying entanglement classes in large-scale quantum systems with numerous subsystems.\\

\begin{figure*}[!t]
    \centering
    \resizebox{14cm}{4cm}{\includegraphics{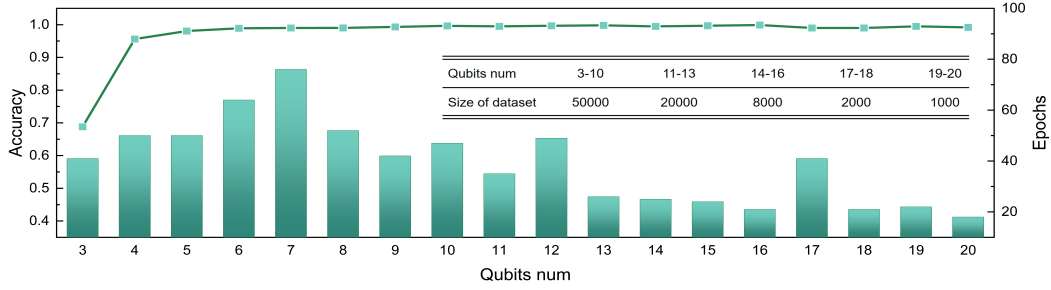}}
    \caption*{Fig.5 The numerical results for the classification between GHZ-class and linear cluster-class in 3-20-qubit systems.  The dot-line plot corresponds to the left vertical axis, representing the classification accuracy of entanglement classes; the histogram corresponds to the right vertical axis, denoting the minimum number of training epochs required for convergence. The size of the dataset corresponding to the number of qubits is shown in the insert panel.}\label{Fig5}
\end{figure*}

To further evaluate the robustness of our method against noise, we perform reliability tests by introducing white noise into the quantum states in accordance with Eq.(2). Specifically, we generate noisy  GHZ-class states and W-class states with fidelities spanning from 0.7 to 0.9 and assess the model's classification performance on these states. The results, as illustrated in Fig.6, demonstrate that our method retains high classification accuracy even in the presence of non-negligible noise. For instance, the model achieves a commendable accuracy on the noisy states when the fidelity is 0.9, with only a slight decrease compared to the performance on states without noise. When the fidelity is 0.7, the accuracy is still acceptable. These findings underscore the robustness and reliability of our neural network-based approach, confirming its applicability to multipartite entanglement classification in practical scenarios where noise is unavoidable.\\

\begin{figure*}[!t]
    \centering
    \resizebox{14cm}{4cm}{\includegraphics{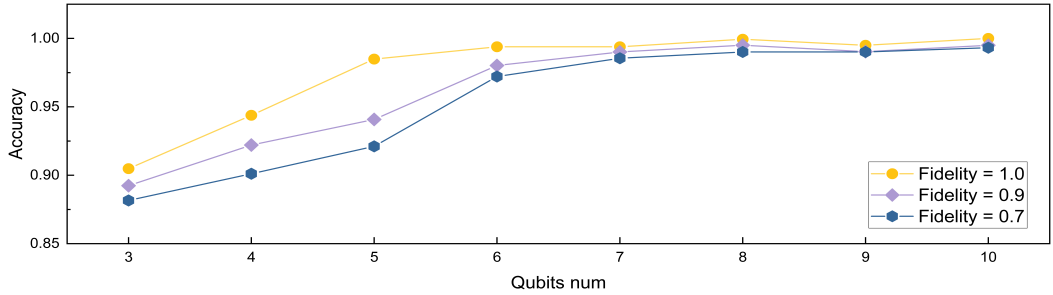}}
    \caption*{Fig.6 The numerical results for the classification for the states with noises.}\label{Fig6}
\end{figure*}

\section{Conclusion}
In this paper, we address the key challenge of multipartite entanglement classification—a task that is increasingly demanding for large-scale quantum systems and faces notable difficulties—by proposing a novel neural network-based classification method guided by the multipartite conditional quantum uncertainty relation. Specifically, we adopt the conditional variance derived from the variance-based quantum-control-assisted multipartite uncertainty relation and take the conditional uncertainty in the variance form as the input features for the neural network, abandoning the traditional density matrix-based feature extraction that suffers from poor scalability for large-scale systems.\\ 

Using the Origin Quantum platform, we adopt the quantum circuit method to generate a total of 1,890,000 quantum states in 3-20-qubit systems, including pure states, mixed states and noisy states, thereby constructing a dedicated dataset for model training and testing. Leveraging the standard MLP as the model architecture, we show that the proposed method is effective, scalable and robust in distinguishing typical multipartite entangled classes, including GHZ-class, W-class and linear cluster-class states, which are SLOCC-distinct. Our method demonstrates high performance for both pure states and mixed states. Even in the presence of non-negligible noise (fidelity ranging from 0.7 to 0.9), the method still maintains a high classification accuracy, with only a slight performance reduction compared to pure state scenario, which confirms its strong anti-white-noise robustness and practical applicability in realistic quantum circuit scenarios where noise is inevitable.\\

The key contribution this work makes to the field of quantum entanglement classification and the integration of quantum physics with machine learning is that it innovatively extracts physical features of entangled states based on quantum uncertainty relations. This strategy enables the model to better capture the intrinsic characteristics of different entanglement classes, and substantially reduces both the measurement consumption and the computational cost by avoiding the adoption of density matrices. This approach represents a brand new avenue for identifying entanglement classes in large-scale quantum systems with numerous subsystems.\\

Despite these achievements, several limitations should be noted. First, the classification accuracy for mixed states, though still high, consistently lags behind that for pure states, which suggests that additional physical features or network architectures may be needed. Second, the method has been evaluated solely via classical numerical simulation; experimental validation on real quantum processors, where device-specific noise and measurement errors are present, is a crucial next step. Third, the numerical validation presented here is based on states generated by random LU transformations. Since LU transformations form a subgroup of SLOCC operations, the generated states remain within the intended SLOCC classes, but the present dataset does not cover states generated by general non-unitary SLOCC operations. Extending the validation to non-unitary SLOCC transformations, which would require more involved numerical and experimental techniques, is an important direction for future work.\\

Beyond the benchmark demonstrations reported here, we highlight several immediate application pathways such as real-time entanglement resource verification or in-situ entanglement monitoring in quantum networks. By bridging fundamental uncertainty principles with neural network design, our work lays a foundation for a new generation of physically interpretable, resource-efficient artificial intelligence tools for quantum information science.\\

\section*{Acknowledgments}
This work was supported by the National Natural Science Foundation of China (NSFC) under Grants No. 12474353.

\section*{Author contributions}
Xiao Zheng and Guo-Feng Zhang conceived and designed the research. Qiyi Li performed analysis and wrote the manuscript with preparing all the figures. All authors discussed the results and revised the manuscript.

\section*{Data availability}
The data that support the findings of this article are available from the corresponding author upon reasonable request.

\end{document}